\documentclass[11pt]{article}

\usepackage[english]{babel}

\usepackage[a4paper,top=2cm,bottom=2cm,left=3cm,right=3cm,marginparwidth=1.75cm]{geometry}

\usepackage{amssymb}
\usepackage{siunitx}
\PassOptionsToPackage{hyphens}{url}
\usepackage[hidelinks]{hyperref}
\usepackage{cleveref}
\usepackage[utf8]{inputenc}
\usepackage{csquotes}
\usepackage{booktabs}
\usepackage{longtable}
\usepackage{adjustbox}
\usepackage{array}
\usepackage{url}
\usepackage{titlesec}
\usepackage{authblk}
\usepackage{xcolor} % Load the xcolor package for color options
\usepackage{siunitx} 
\usepackage{subcaption} % サブキャプションのサポート
\usepackage{wrapfig}
\usepackage[english]{babel}
\crefformat{figure}{#2Figure~#1#3}
\Crefformat{figure}{#2Figure~#1#3}
\crefformat{table}{#2Table~#1#3}
\Crefformat{table}{#2Table~#1#3}
\crefformat{section}{#2Section~#1#3}
\Crefformat{section}{#2Section~#1#3}

\author{Naoki Otani* for the NINJA Collaboration}

\affil{*Kyoto University}

\title{The Preparation Status and Plan for the Next Physics Run of the NINJA Experiment \\\resizebox{\textwidth}{!}{``Contribution to the 25th International Workshop on Neutrinos from Accelerators''}}

\begin{document}
\maketitle

\begin{abstract}
The NINJA experiment aims to precisely measure neutrino-nucleus interactions using a nuclear emulsion detector to reduce systematic errors in neutrino oscillation experiments. 
The nuclear emulsion has a sub-micron positional resolution, enabling the detection of low-momentum charged particles such as protons with a threshold of 200 MeV/c. 
In the NINJA experiment, a muon detector placed downstream of the emulsion detector is used to identify muons from $\nu_{\mu}$ charged-current interactions.
The majority of the tracks accumulated in the nuclear emulsion are from cosmic rays.
Although the emulsion detector provides highly accurate positional information, it lacks timing information.
Therefore, the positional resolution of the muon detector is not enough to identify neutrino interaction tracks that match between the muon detector and the emulsion detector from the enormous background of cosmic rays recorded in the emulsion detector.
To address this, a scintillation tracker is used to provide both timing and positional information for the tracks. 

The NINJA experiment is planning a third physics run with about 130 kg water target from the autumn of 2025 to the spring of 2026. 
Since the target mass is larger than previous runs, a larger scintillation tracker covering 1.3~m~$\times$~1.4~m is needed.
We are developing a newly designed scintillation tracker, consisting of a monolithic plastic scintillator plane including scatterers. 

In this paper, we will show the preparation status and plan for the next physics run, focusing particularly on the development of the new scintillation tracker.
\end{abstract}
% % \linenumbers

% \section*{Template Overview}
% \textcolor{blue}{This \LaTeX template is designed to incorporate the specific Harvard citation style defined by Emerald Publishing, along with section styles and other adjustments required for submission in Construction Innovation journal.
% Users should adapt the styling to align with the guidelines of the journal to which they are submitting their article. Make sure to remove the authors' names and acknowledgements if you are submitting an anonymous file for double-blind peer review.}

\section{Introduction}
The uncertainty of neutrino-nucleus interaction is one of the major systematic errors in neutrino oscillation experiments, such as the T2K experiment\cite{t2ksystematic}. 
It is crucial to comprehend the 2p2h (2 particle - 2 hole) interaction in particular.
In the T2K experiment, the neutrino energy is reconstructed from charged-lepton kinematics under the assumption of CCQE (Charged Current Quasi Elastic) scattering being the dominant mode of interaction.
If the contamination rate of 2p2h interactions is not accurately known, the reconstructed energy is biased.
However, there have not been sufficient measurements of protons from 2p2h interactions so far because the detection of low-momentum protons produced from 2p2h interactions is difficult.
Consequently, this leads to significant uncertainties in 2p2h interaction models.
% ニュートリノ反応の不定性はT2K実験での系統誤差の主な原因
% 特に2p2h反応の理解が重要である．
% なぜなら2p2h反応から出てくる核子は低運動量のためにCCQEと区別することが難しく、2p2h反応の混入率が正確にわかっていないと再構成されたエネルギーにバイアスがかかる
% ニュートリノ振動確率はエネルギーに依存するのでエネルギーを正確に再構成することが重要

\section{NINJA Experiment}

The goal of NINJA (Neutrino Interaction research with Nuclear emulsion and J-PARC Accelerator) experiment is to precisely measure sub-multi~GeV neutrino interactions using a nuclear emulsion detector. 
The nuclear emulsion, with its sub-micron positional resolution, allows for detecting low-momentum charged particles such as protons with a threshold of $200 ~\mathrm{MeV}/c$.
The NINJA experiment is capable of covering almost the entire momentum range of protons from 2p2h interactions.

So far, the NINJA experiment has conducted two physics runs in J-PARC (Japan Proton Accelerator Research Complex) and acquired data corresponding to $7.7\times 10^{20}~\mathrm{POT}$ (Protons on Target) with a 75 kg water target.
The NINJA experiment is planning a third physics run with a 130 kg water target from the autumn of 2025 to the spring of 2026.

\section{Detectors}

\begin{figure}[htbp]
    \centering
    % 左側の図
    \begin{minipage}{0.4\textwidth}
        \centering
        \includegraphics[width=\linewidth]{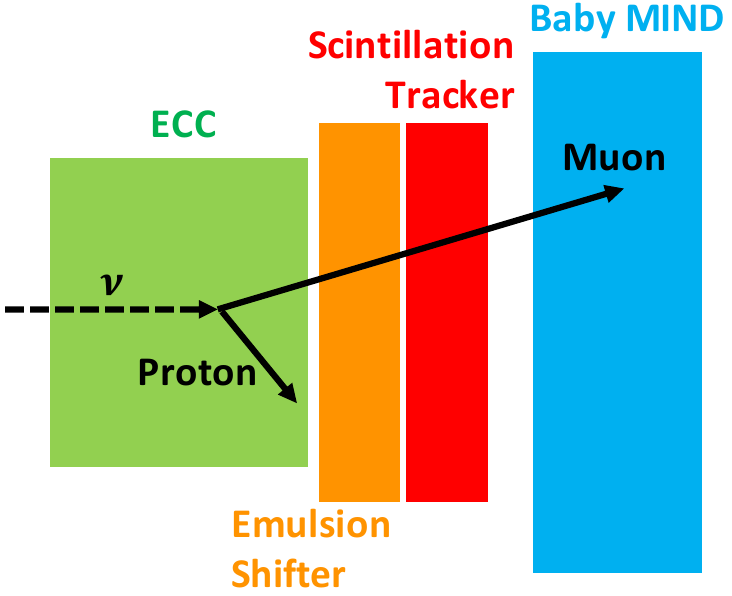}
        \caption{The setup of the detectors}
        \label{fig:ninjasetup}
    \end{minipage}
    \hfill
    % 右側の図
    \begin{minipage}{0.5\textwidth}
        \centering
        \includegraphics[width=\linewidth]{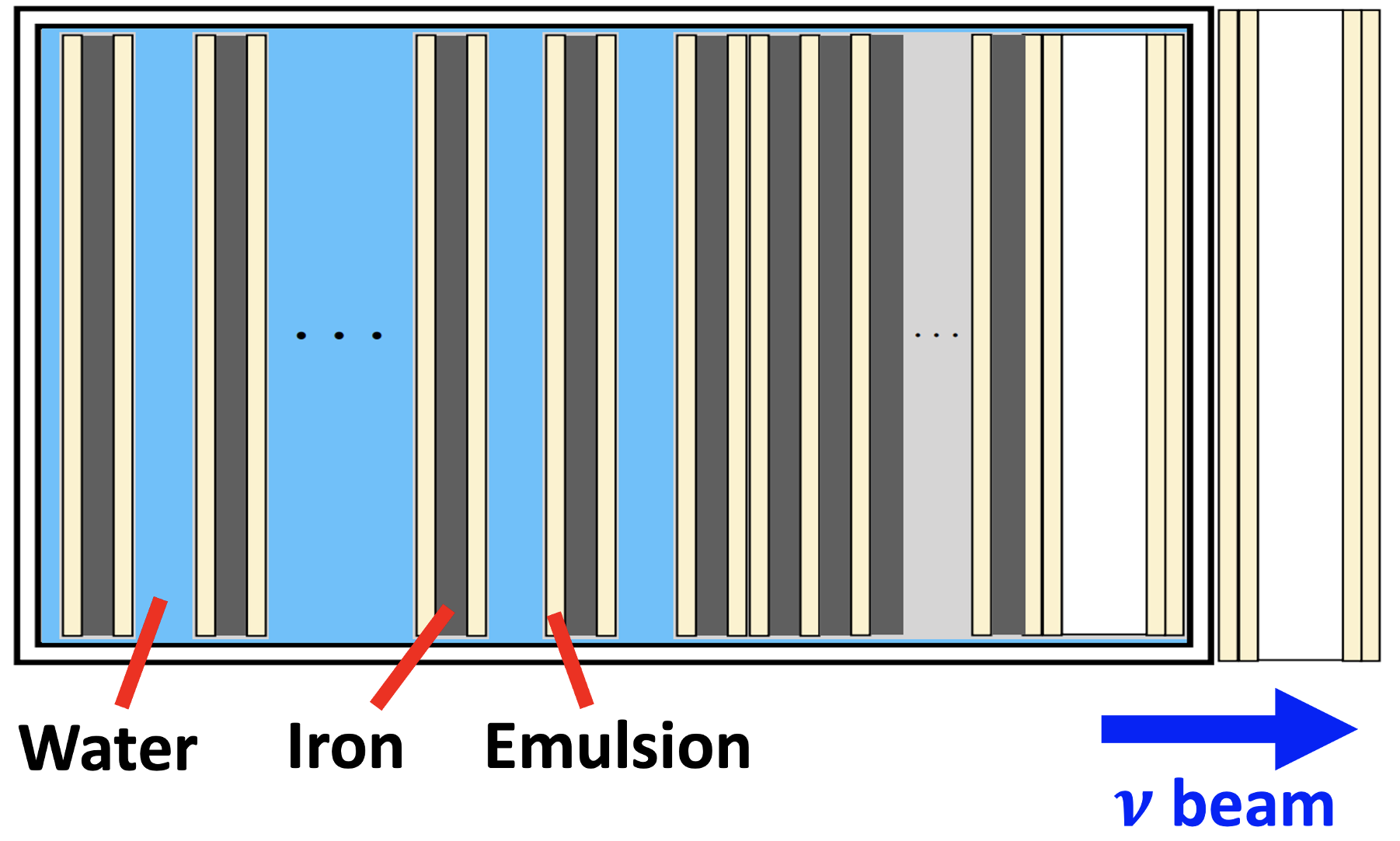}
        \caption{The structure of the ECC}
        \label{fig:eccstructure}
    \end{minipage}
\end{figure}

Figure \ref{fig:ninjasetup} shows the setup of the detectors in the NINJA experiment.
The main emulsion detectors called ECCs (Emulsion Cloud Chambers) are placed at the most upstream. 
The T2K near detectors include a muon range detector at the farthest downstream called Baby MIND (Magnetized Iron Neutrino Detector)\cite{babymind}.
% A muon range detector called Baby MIND (Magnetized Iron Neutrino Detector)\cite{babymind}, which is a part of the T2K near detectors, is placed at the most downstream. 
Baby MIND has a sandwich structure consisting of magnetized iron plates and plastic scintillator bars, providing the kinematics of muons going out of the ECCs.

The majority of the tracks accumulated in the ECCs are from cosmic rays, and although the nuclear emulsion provides a good positional resolution, it lacks timing information.
Consequently, the positional resolution of Baby MIND is insufficient to identify neutrino interaction tracks that match between Baby MIND and the ECCs from the enormous background of cosmic rays recorded in the ECCs.
To address this, two types of detectors are placed between the ECCs and Baby MIND.
One is an emulsion shifter, consisting of moving and fixed emulsion films, which offers both highly accurate position information and approximate timing information.
The other is a scintillation tracker, composed of a plastic scintillator, which provides beam timing information and better position information than Baby MIND.

\subsection{ECC (Emulsion Cloud Chamber)}
An ECC is composed of alternating layers of tracking units and water targets as shown in Figure \ref{fig:eccstructure}. 
Each tracking unit has two 350\,\si{\micro\metre}-thick emulsion films on both sides of a 500~\si{\micro\metre}-thick iron plate, which is used for a momentum measurement.
The tracks of charged particles from neutrino interactions in the water target are recorded in the emulsion films.

We installed $3\times3$ ECCs in the earlier runs, but we intend to install $4\times4$ ECCs in the upcoming run.
The production of the ECCs will start in the spring of 2025.

\subsection{Emulsion Shifter}
Since the target mass will be larger in the next run, we will use a new emulsion shifter with a size of 1.2~m~$\times$~1.4~m, larger than the 1.0~m~$\times$~1.0~m emulsion shifter used in the previous runs.
The emulsion shifter consists of two moving walls and one fixed wall, on which emulsion films are mounted.
Each moving wall moves 1.2 mm every 3~minutes and 12~hours, respectively.
Timing information with a 3-minute time resolution, provided by the track connection between the moving walls and the fixed walls, is better than 4-hour time resolution in earlier runs.

The construction of the new emulsion shifter has already been conducted in Nagoya University and the operation test is ongoing.

\subsection{Scintillation Tracker}
\begin{figure}[htbp]
    \centering
    % 左側の図
    \begin{minipage}{0.5\textwidth}
        \centering
        \includegraphics[width=\linewidth]{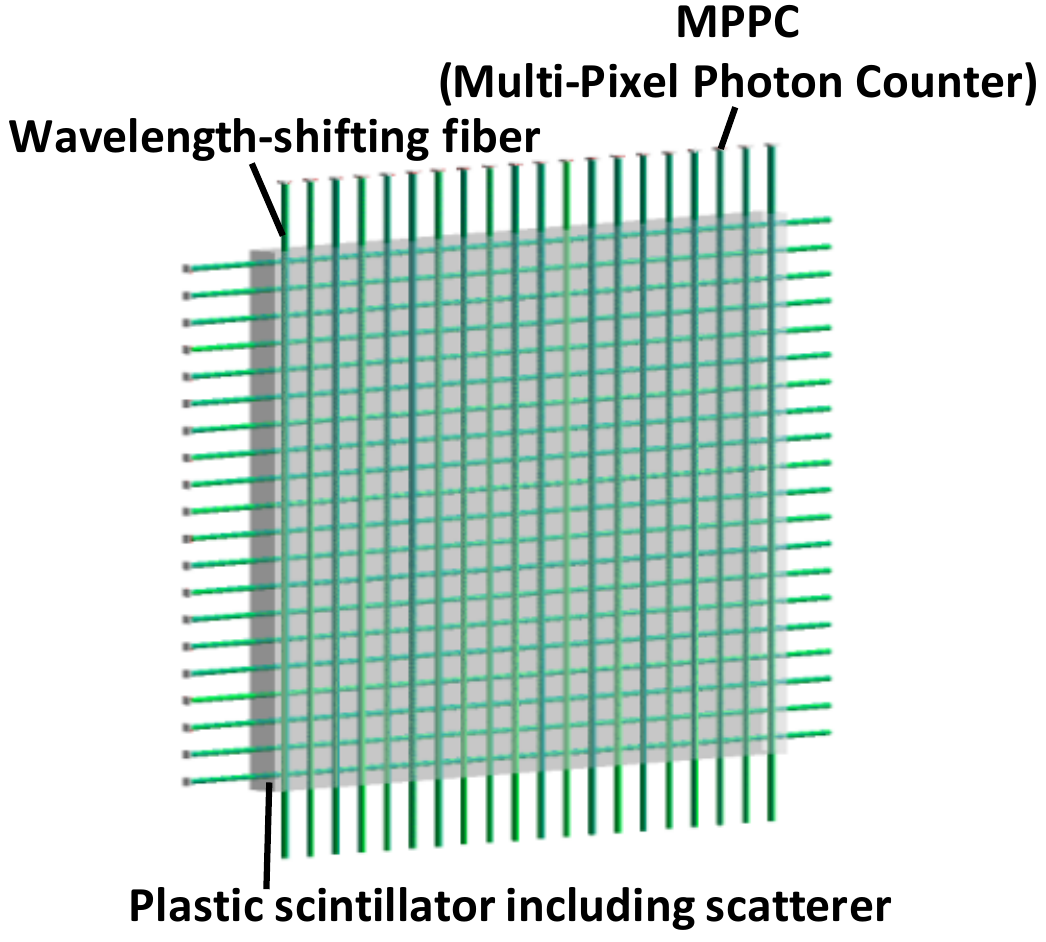}
        \caption{The design of a new scintillation tracker}
        \label{fig:trackerdesign}
    \end{minipage}
    \hfill
    % 右側の図
    \begin{minipage}{0.47\textwidth}
        \centering
        \includegraphics[width=\linewidth]{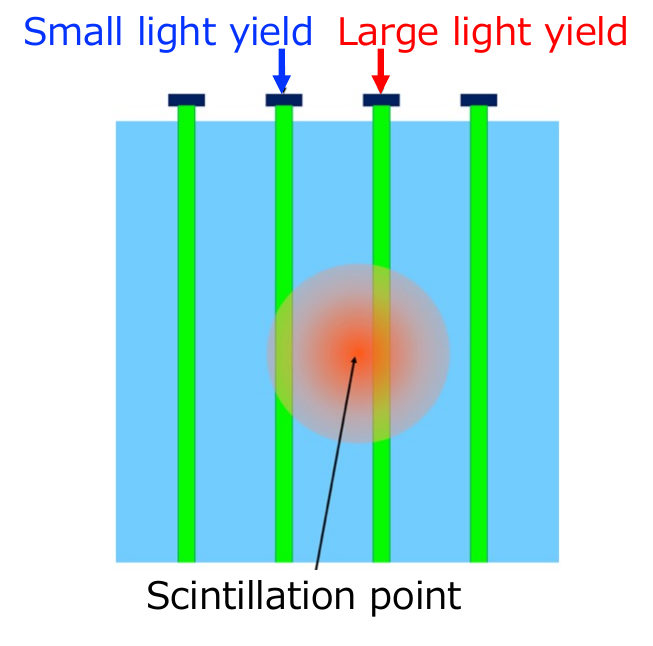}
        \caption{The mechanism of position reconstruction}
        \label{fig:trackermechanism}
    \end{minipage}
\end{figure}
In the next run, due to the increased target mass, we need a new scintillation tracker with a size of 1.3~m~$\times$~1.4~m, larger than the 1.0~m~$\times$~1.0~m tracker used in the previous runs. 
The scintillation tracker used in the previous runs is composed of 248 scintillator bars.
Expanding this design results in an increased number of readout channels.
To achieve a larger size without increasing the number of readout channels, we are developing a newly designed scintillation tracker.

\begin{wrapfigure}[14]{r}[0pt]{6.5cm}
\begin{center}
		\raisebox{-42mm}[0pt][42mm]{\includegraphics[width=7cm]{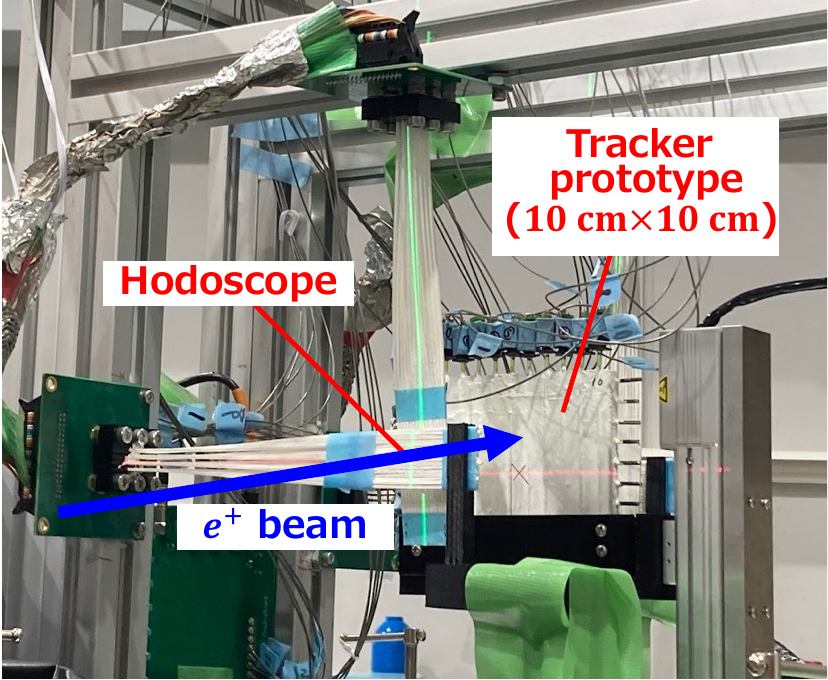}}
	\caption{The setup of the beam test} 
        \label{fig:beamtest}
\end{center}
\end{wrapfigure}
The new scintillation tracker consists of a monolithic plastic scintillator plane including scatterers as shown in Figure \ref{fig:trackerdesign}.
The scintillation light is read out by wavelength-shifting fibers and MPPCs (Multi-Pixel Photon Counters) at 10~mm intervals.
The mechanism of position reconstruction is shown in Figure \ref{fig:trackermechanism}.
The light yield becomes larger for channels closer to a scintillation point.
The position of a charged particle passing through the scintillation tracker is reconstructed by using the light yield balance information.

% \begin{figure}[ht]
%  \centering
%  \makebox[\textwidth][c]{\includegraphics[width=0.4\textwidth]{Figures/beamtest_setup.pdf}}%
%  \caption{The setup of the beam test}
%  \label{fig:beamtest}
% \end{figure}

We evaluated the performance of the new scintillation tracker using the tracker prototypes and the positron beam at RARiS, Tohoku University, in July 2024.
Figure \ref{fig:beamtest} shows the setup of the beam test.
The tracker prototype with a size of 10~cm~$\times$~10~cm was placed between the hodoscopes.
The hodoscopes, composed of 1.7~mm width scintillation fibers, were used to determine the beam position.
The results show that the positional resolution is 1.44~mm when a charged particle enters perpendicularly and 1.84~mm when the incident angle is 45 degrees.
These results demonstrate that the new scintillation tracker can determine the position of a charged particle with much better precision than the required positional resolution of 14.6~mm.

The construction of the actual tracker will start around February 2025.

\section{Prospects of the Physics Results}

\begin{figure}[htbp]
 \centering
 \makebox[\textwidth][c]{\includegraphics[width=0.85\textwidth]{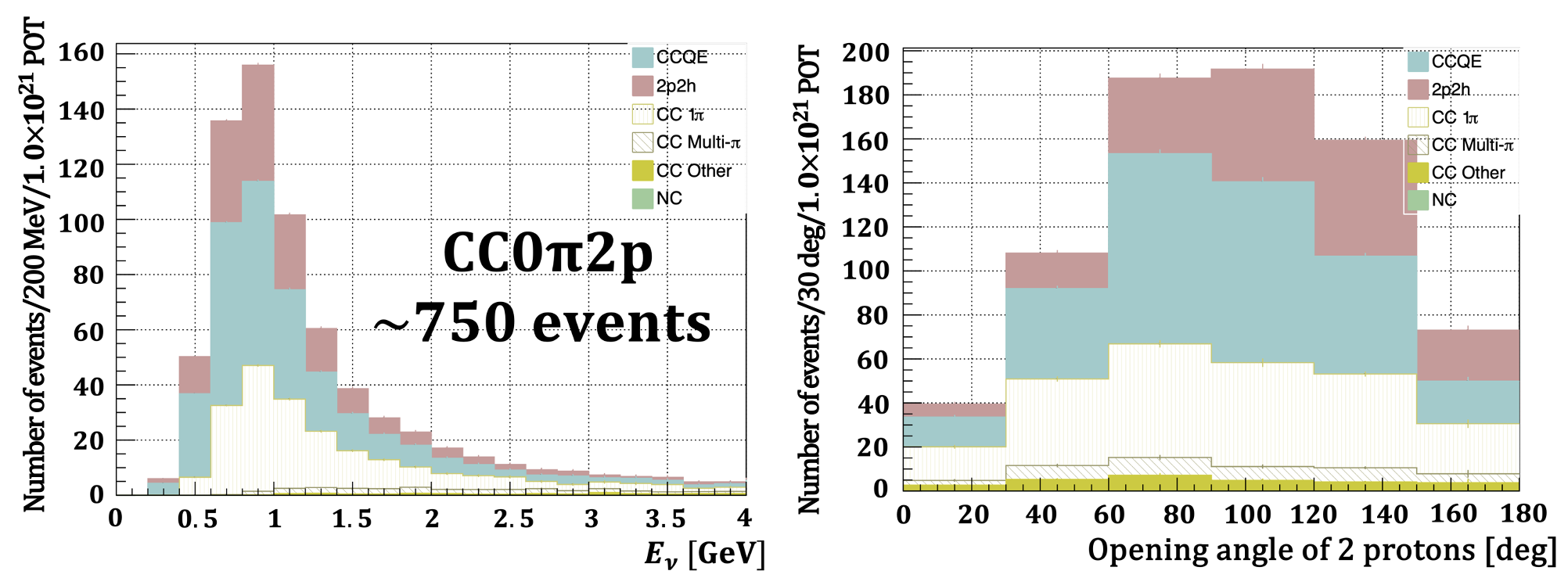}}%
 \caption{The expected number of CC0$\pi$2p events (left) and the distribution of the opening angle of two protons in CC0$\pi$2p (right) in all physics runs ($1.0\times 10^{21}$ POT)}
 \label{fig:cc0pi2p}
\end{figure}

Figure \ref{fig:cc0pi2p} shows the expected number of CC0$\pi$2p events and the distribution of the opening angle of two protons in  CC0$\pi$2p in all physics runs with $1.0\times 10^{21}$ POT beam exposure.
The interactions are simulated by NEUT 5.4.0\cite{neut}, and the Nieves et al.\cite{2p2hmodel} model is applied to the 2p2h interactions.
We will observe approximately 750 CC0$\pi$2p events out of roughly 5480 CC events in all physics runs.
2p2h interactions tend to have a large opening angle of two protons. 
This characteristic will be the key to constraining the 2p2h interaction model.

% \begin{figure}[htbp]
%   \centering
%   \begin{subfigure}{0.47\textwidth}
%     \centering
%     \includegraphics[width=\linewidth]{Figures/cc0pi2pneut.png}
%     \caption{Caption for image 1}
%     \label{fig:image1}
%   \end{subfigure}
%   \hfill
%   \begin{subfigure}{0.47\textwidth}
%     \centering
%     \includegraphics[width=\linewidth]{Figures/openingangleneut.png}
%     \caption{Caption for image 2}
%     \label{fig:image2}
%   \end{subfigure}
%   \caption{A set of images placed side by side.}
%   \label{fig:side_by_side}
% \end{figure}

\section{Conclusion}
In conclusion, the NINJA experiment aims for a precise measurement of neutrino
interactions using a nuclear emulsion to reduce the systematic errors in neutrino oscillation experiments.
The NINJA experiment will start a third physics run with a 130~kg water target in the autumn of 2025.
Every detector, including the ongoing development of the new scintillation tracker, is in the process of being prepared.
The result of the beam test shows that the new scintillation tracker can reconstruct positions with much better resolution than required.
We aim to constrain the neutrino interaction models using the kinematics information such as the opening angle of two protons in CC0$\pi$2p.

\hfill

\textit{Acknowledgments} \\Part of this study was performed using facilities of Research Center for Accelerator and Radioisotope Science (RARiS), Tohoku University. We thank RARiS for the allocation of beamtime.

%\break
% \printbibliography

\end{document}